\documentclass{wscpaperproc}
\usepackage{latexsym}
\usepackage{graphicx}
\usepackage{mathptmx}
\usepackage[T1]{fontenc}

\usepackage{caption} 
\usepackage{amsmath}
\usepackage{amsfonts}
\usepackage{amssymb}
\usepackage{amsbsy}
\usepackage{amsthm}

\usepackage[pdftex,colorlinks=true,urlcolor=blue,citecolor=black,anchorcolor=black,linkcolor=black]{hyperref}

\usepackage{xcolor}

\newcommand{\secref}[1]{Section~\ref{#1}}

\newtheoremstyle{wsc}
{3pt}
{3pt}
{}
{}
{\bf}
{}
{.5em}
{}

\theoremstyle{wsc}

\begin{document}

%
%

\pagestyle{fancyplain}

\thispagestyle{plain}
\firstPageHead{}

\chead{\fancyplain{}{\itshape Giabbanelli, Beverley, David, and Tolk}}

\rhead{}
\cfoot{}
\renewcommand{\headrulewidth}{0pt} 

\makeatletter
\let\@internalcite\cite
\def\cite{\def\@citeseppen{-1000}%
    \def\@cite##1##2{(##1\if@tempswa , ##2\fi)}%
    \def\citeauthoryear##1##2##3{##1 ##3}\@internalcite}
\def\citeNP{\def\@citeseppen{-1000}%
    \def\@cite##1##2{##1\if@tempswa , ##2\fi}%
    \def\citeauthoryear##1##2##3{##1 ##3}\@internalcite}
\def\citeN{\def\@citeseppen{-1000}%
    \def\@cite##1##2{##1\if@tempswa, ##2)\else{}\fi}%
    \def\citeauthoryear##1##2##3{##1 (##3)}\@citedata}
\def\citeA{\def\@citeseppen{-1000}%
    \def\@cite##1##2{(##1\if@tempswa , ##2\fi)}%
    \def\citeauthoryear##1##2##3{##1}\@internalcite}
\def\citeANP{\def\@citeseppen{-1000}%
    \def\@cite##1##2{##1\if@tempswa , ##2\fi}%
    \def\citeauthoryear##1##2##3{##1}\@internalcite}
\def\shortcite{\def\@citeseppen{-1000}%
    \def\@cite##1##2{(##1\if@tempswa , ##2\fi)}%
    \def\citeauthoryear##1##2##3{##2 ##3}\@internalcite}
\def\shortciteNP{\def\@citeseppen{-1000}%
    \def\@cite##1##2{##1\if@tempswa , ##2\fi}%
    \def\citeauthoryear##1##2##3{##2 ##3}\@internalcite}
\def\shortciteN{\def\@citeseppen{-1000}%
    \def\@cite##1##2{##1\if@tempswa, ##2\else{}\fi}%
    \def\citeauthoryear##1##2##3{##2 (##3)}\@citedata}
\def\shortciteA{\def\@citeseppen{-1000}%
    \def\@cite##1##2{(##1\if@tempswa , ##2\fi)}%
    \def\citeauthoryear##1##2##3{##2}\@internalcite}
\def\shortciteANP{\def\@citeseppen{-1000}%
    \def\@cite##1##2{##1\if@tempswa , ##2\fi}%
    \def\citeauthoryear##1##2##3{##2}\@internalcite}
\def\citeyear{\def\@citeseppen{-1000}%
    \def\@cite##1##2{(##1\if@tempswa , ##2\fi)}%
    \def\citeauthoryear##1##2##3{##3}\@citedata}
\def\citeyearNP{\def\@citeseppen{-1000}%
    \def\@cite##1##2{##1\if@tempswa , ##2\fi}%
    \def\citeauthoryear##1##2##3{##3}\@citedata}
%
%
%
\def\@citedata{%
    \@ifnextchar [{\@tempswatrue\@citedatax}%
                  {\@tempswafalse\@citedatax[]}%
}

\def\@citedatax[#1]#2{%
\if@filesw\immediate\write\@auxout{\string\citation{#2}}\fi%
  \def\@citea{}\@cite{\@for\@citeb:=#2\do%
    {\@citea\def\@citea{, }\@ifundefined
       {b@\@citeb}{{\bf ?}%
       \@warning{Citation `\@citeb' on page \thepage \space undefined}}%
{\csname b@\@citeb\endcsname}}}{#1}}%

%
\def\@citex[#1]#2{%
\if@filesw\immediate\write\@auxout{\string\citation{#2}}\fi%
  \def\@citea{}\@cite{\@for\@citeb:=#2\do%
    {\@citea\def\@citea{; }\@ifundefined
       {b@\@citeb}{{\bf ?}%
       \@warning{Citation `\@citeb' on page \thepage \space undefined}}%
{\csname b@\@citeb\endcsname}}}{#1}}%

%
\def\@biblabel#1{}
\makeatother



\newdimen\bibindent
\bibindent=0.0em
\def\thebibliography#1{\section*{\refname}\list
   {}{\settowidth\labelwidth{[#1]}
   \leftmargin\parindent
   \itemindent -\parindent
   \listparindent \itemindent
   \itemsep 0pt
   \parsep 0pt}
   \def\newblock{}
   \sloppy
   \sfcode`\.=1000\relax}


\setlength{\baselineskip}{12.7pt}

\title{FROM OVER-RELIANCE TO SMART INTEGRATION: USING LARGE-LANGUAGE MODELS AS TRANSLATORS BETWEEN SPECIALIZED MODELING AND SIMULATION TOOLS}

\author{\begin{center}Philippe J. Giabbanelli\textsuperscript{1}, John Beverley\textsuperscript{2}, Istvan David\textsuperscript{3}, and Andreas Tolk\textsuperscript{4}\\
[11pt]
\textsuperscript{1}Virginia Modeling, Analysis, and Simulation Center (VMASC), Old Dominion University, VA, USA\\
\textsuperscript{2}Dept.~of Philosophy, University at Buffalo, Buffalo, NY, USA\\
\textsuperscript{3}McMaster Centre for Software Certification (McSCert), McMaster University, Hamilton, Canada\\
\textsuperscript{4}The MITRE Corporation, Charlottesville, VA, USA\end{center}
}

\maketitle

\vspace{-12pt}

\section*{ABSTRACT}
Large Language Models (LLMs) offer transformative potential for Modeling \& Simulation (M\&S) through natural language interfaces that simplify workflows. However, over-reliance risks compromising quality due to ambiguities, logical shortcuts, and hallucinations. This paper advocates integrating LLMs as middleware or translators between specialized tools to mitigate complexity in M\&S tasks. Acting as translators, LLMs can enhance interoperability across multi-formalism, multi-semantics, and multi-paradigm systems. We address two key challenges: identifying appropriate languages and tools for  modeling and simulation tasks, and developing efficient software architectures that integrate LLMs without performance bottlenecks.
To this end, the paper explores LLM-mediated workflows, emphasizes structured tool integration, and recommends Low-Rank Adaptation-based architectures for efficient task-specific adaptations. This approach ensures LLMs complement rather than replace specialized tools, fostering high-quality, reliable M\&S processes.

\section{INTRODUCTION}
\label{sec:intro}
Large Language Models (LLMs) offer the temptation of a one-stop-shop by handling seemingly any Modeling \& Simulation (M\&S) task, at the easily satisfiable condition of expressing the task through natural language and/or through images. This vastly simplifies the workflow of modelers, by forging the illusion that it is no longer necessary to navigate different tools and data representations. As a result, LLMs often are adopted due to the perceived \textit{convenience} of bypassing the strenuous steps of de- and re-composing complex problems for modeling and analysis purposes. Such convenient shortcuts conceal accidental and essential complexity of the problem at hand and therefore, excessive reliance on LLMs may have detrimental effects for M\&S research.

Research quality is negatively impacted when the LLM output is \textit{suboptimal} relative to specialized tools or expertise. Modelers may deem the LLM outputs sufficiently accurate if they are not aware that a better, specialized solution exists. For example, the LLM may be asked to think about whether changes in the input of a simulation model would create a desired effect. Directly using a LLM may result in logical shortcuts or hallucinations. However, LLMs are proficient at translating such queries into symbolic representations (e.g., first order logic) and they can call a symbolic solver to guarantee logical coherence during reasoning \shortcite{lam2024closer}. 


In this paper, we focus on leveraging the convenience and simplicity of using LLMs without sacrificing quality or losing the benefits of existing investments in specialized tools.
\textbf{We posit that LLMs should be used to mitigate accidental complexity, so that the human can focus on the essential complexity of the problem at hand.}
Since accidental complexity in M\&S research typically stems from multi-formalism, multi-semantics, multi-paradigm modeling, and the supporting heterogeneous toolchains, we propose that modelers should leverage LLMs as a ``glue'' or middleware ``translator'' between specialized tools and expertise. That is, instead of aiming to perform any task solely \textit{within} a LLM, the emphasis should be on placing LLMs \textit{between} modeling and simulation systems to make them more easily interoperable.

The realization that LLMs need to be integrated with specialized tools is gaining traction in M\&S research~\shortcite{ZHANG2024114788} and echoes a broader interest in positioning LLMs as middleware instead of the sole solution~\shortcite{lehmann2024towards,calvetcan}. In this paper, our main contribution is to address the two core questions raised by the opportunities and challenges of integrating LLMs with specialized tools to accomplish modeling and simulation tasks:
\begin{itemize}
    \item[\textbf{Q1}] If the LLM is to act as a translator, then which \textit{language} should it translate into and which \textit{tool} should it call depending on the M\&S task? 
    \item[\textbf{Q2}] Which software \textit{architecture} could efficiently integrate LLM as a middleware to support the interoperability of specialized M\&S tools without creating performance bottlenecks?
\end{itemize}

The remainder of this paper is organized as follows. \secref{sec:background} succinctly demonstrates across fields that LLMs have been used for their convenience, occasionally at the expense of quality. This evidence base, across applications, motivates our paper by showing that the understandable need for convenience from M\&S practitioners should not result in an exclusive reliance on LLMs, but rather by employing LLMs in well-defined roles such as a middleware between specialized tools. In \secref{sec:conceptual}, we identify languages and tools to perform  modeling tasks (\textbf{Q1}), ranging from building models (e.g., by merging or from a corpus) to validating and explaining their structure. Similarly, \secref{sec:simulation} covers the languages and tools for simulation tasks (\textbf{Q1}). In \secref{sec:architectures}, we review the architectures available (\textbf{Q2}) and propose one solution that has yet to be employed for M\&S but has a strong potential for high performances in both training and execution. Finally, \secref{sec:discussion} examines key issues in using LLMs as middleware, including \textit{trust}, the \textit{reciprocal benefits} between LLMs and specialized tools, and the need for extensive \textit{benchmarks}.

\section{BACKGROUND}
\label{sec:background}
\subsection{Evidence for the Convenient Use of LLMs at the Detriment of Quality Across Applications}
\label{sec:convenience}
An already infamous Forbes report from 2023 illustrates that LLMs are used as a `Swiss-Army Knife' for diverse tasks because they are easy to use, despite the existence of tools that might produce higher-quality outputs. This report describes a lawyer who used ChatGPT to draft and submit a defense brief that contained fictitious case precedents~\cite{novak2023lawyer}. While the efficacy and accuracy of LLMs used in the legal domain has significantly increased since 2023, the precision demanded in legal discourse within and across languages raises barriers just as significant~\shortcite{padiu2024extent}. As a result, the ease of using LLMs in such cases is outweighed by the risk of legal missteps that are more likely to be avoided by legal experts.  

In the clinical domain, \shortciteN{brown2024not} compared traditional machine learning (ML) methods with LLMs (GPT-3.5 and GPT-4) for prediction tasks using electronic health records. Their study found that while LLMs are attractive for their simplicity and ease of deployment, traditional ML approaches—such as gradient-boosted trees—achieved significantly higher predictive performance and calibration. From another direction, \shortciteN{van2024adapted} reported that although LLMs can produce clinical summaries that, in many cases, are judged comparable to those generated by experts, the models still exhibit safety concerns such as hallucinatory content, which specialized summarization systems or curation methods employed by experts are better at avoiding. This is not to say continued research will not ultimately address LLM limitations in this domain or others. Rather, it is to highlight that despite the impressive applications of LLMs in this area~\shortcite{yu2024large,zheng2024large}, general applicability should be viewed with caution. 

In the area of code generation, \citeN{tang2024towards} and \citeN{basic2024large} highlight that LLM-generated code, while easy to obtain, often contains bugs and security vulnerabilities that require manual, and often time-consuming, efforts to untangle. In a reverse engineering context, \citeN{pordanesh2024exploring} demonstrated that GPT-4 can offer broad insights when analyzing decompiled code, but it frequently struggles with the intricacies of complex code structures—leading to superficial analyses compared to established reverse engineering tools, such as the disassembler and decompiled Binary Ninja~\cite{binaryninja} and the Ghidra suite~\cite{ghidra}.

\subsection{Can LLMs address the Elusiveness of Simulation Interoperability}

Simulation interoperability has been pursued through various research and engineering efforts for over three decades \cite{tolk2024conceptual}. Many approaches have fallen short because they treated simulation systems as mere software solutions. In reality, simulations implement models that are developed for specific purposes: to address particular research questions or support distinct tasks. This means that representations of simulated entities are rarely identical across different systems -- their scope and structure naturally differ. This underlying model diversity is the fundamental reason why simulation interoperability remains elusive: technical standards that interconnect simulations cannot resolve conceptual differences between models.

LLMs offer promising potential to address this challenge. They can help bridge both referential differences (what information a simulation provides) and methodological differences (how this information is structured). Frydenlund et al.~\citeyear{frydenlund2024modeler} demonstrated using the same model to create simulations employing different paradigms, achieving varying levels of success. Jackson and Rolf \citeyear{jackson2023natural} utilized LLMs to generate natural language descriptions from simulation data. 

Evidence suggests LLMs have strong translation capabilities. OpenAI's GPT-4 has demonstrated ability to translate between numerous languages according to their technical documentation and public demonstrations, while Google has progressively integrated LLM technology into their translation services, beginning with Neural Machine Translation systems and evolving to include models like T5 and PaLM. Although neither company fully discloses their production architecture details, their capabilities suggest that if LLMs can effectively translate between different linguistic structures and grammars, they should increasingly contribute to addressing the resolution, paradigm, and scope differences that have historically challenged simulation interoperability.

A new research domain are execution-aware LLMs \shortcite{di2025Execution}. While the current articles focusing predominantly on source code generation that does not only consider the static source code but also considers how this source code is executed and what effect such executions have, these ideas clearly point beyond the usual semantic understanding of an artifact towards a pragmatic understanding of how this artifact is used. This new domain has the potential to go beyond source code engineering and address the higher levels of interoperability as requested in \cite{tolk2024conceptual}. However, the research is still in its infancy and the authors are not aware of any simulation-specific applications thereof.

\section{LLMs for modeling}
\label{sec:conceptual}

The flexibility of LLMs often comes at the cost of output precision, formal reasoning capabilities, and domain-specific accuracy. Modeling is a task driven activity to select the scope of the model, as well as the appropriate abstraction level that results in a specification that can be implemented. As such, it involves tasks that range from constructing model specifications (e.g., by merging existing models or extracting models from data) to verifying and explaining model structures. Many such tasks require structured data representations, formal logic, and domain-aware processing, which specialized tools handle more effectively than LLMs. Table 1 summarizes key  modeling tasks and examples of languages and tools that directly support each task out-of-the-box, without requiring custom integration code. 

\begin{table}[htb]
\centering
\caption{Modeling tasks and examples of languages/tools with native out-of-the-box support.}
\footnotesize
\begin{tabular}{|p{4.2cm}|p{4.7cm}|p{5.4cm}|}
\hline
\textbf{Modeling Task} & \textbf{Example Language/Notation} & \textbf{Specialized Tool (Direct Support)} \\
\hline
Merging Models & OWL ontologies & \textit{OntoAligner}, \textit{CoMerger}, \textit{Protégé} with PROMPT plugin \shortcite{protege2004} \\
\hline
Building Model from Corpus & Network text analysis & \textit{AutoMap} – extracts and visualizes concept networks from text \shortcite{automap2013} \\
\hline
Validating Model Structure & First-order logic (Alloy), OWL, UML & \textit{Alloy Analyzer} \shortcite{alloy2006}, HermiT reasoner in \textit{Protégé} \shortcite{hermit2014}, \textit{UModel} \\
\hline
Explaining/Documenting Models & Structured model descriptions & \textit{Enterprise Architect} \shortcite{enterprise2023}, Protégé OWL2NL plugin \\
\hline
Data/Schema Modeling & ER diagrams, UML & \textit{ER/Studio}, \textit{Erwin Data Modeler} \shortcite{erwin2023} \\
\hline
System Modeling and Simulation & Modelica & \textit{Wolfram SystemModeler} \shortcite{systemmodeler2023} \\
\hline
\end{tabular}
\label{tab:conceptual-tools}
\vspace{-0.9em} 
\captionsetup{skip=0pt}
\caption*{\footnotesize \textit{Note: OWL is the Web Ontology Language, UML is the Unified Modeling Language.}}
\end{table}

Table 2 provides an overview of contemporary tools that directly support major  modeling tasks without requiring custom integration. These tools are designed to reduce manual effort and improve modeling reliability across M\&S domains. For example, OntoAligner and CoMerger support semi-automated merging of OWL ontologies with built-in alignment and consistency checks, ideal for collaborative modeling or integrating heterogeneous sources in domains like defense and logistics. Legacy tools like the PROMPT plugin for Protégé remain useful for smaller or more controlled merge tasks. AutoMap enables rapid  model generation from unstructured text by extracting and visualizing concept networks. This is particularly useful in healthcare or intelligence analysis, where narrative data, such as that reflecting patient reports or incident logs, can be converted directly into influence diagrams or causal maps.

\begin{table}[htb]
\centering
\caption{LLM-mediated  modeling workflows: tools, roles, and integration patterns.}
\footnotesize
\begin{tabular}{|p{3.2cm}|p{3.8cm}|p{3.7cm}|p{4cm}|}
\hline
\textbf{Modeling Subtask} & \textbf{Recommended Tool(s)} & \textbf{LLM Role} & \textbf{Integration Notes} \\
\hline
Drafting initial model structure from natural language & GPT-4 + \textit{UModel}, \textit{ER/Studio}, \textit{Protégé} & Translates textual input to structured representation (UML, OWL, ER) & Output should target tool-compatible syntax (e.g., XMI, RDF/XML); tool validates structure \\
\hline
Ontology/model alignment & \textit{OntoAligner}, \textit{CoMerger}, Protégé PROMPT & Suggests initial mappings between terms; explains inconsistencies & Specialized tool performs alignment, consistency check; LLM mediates error resolution \\
\hline
Data schema mapping and reconciliation & \textit{OpenRefine} \shortcite{openrefine2023}, \textit{Neo4j}, \textit{spaCy} & Translates error logs; proposes entity matches & LLM can explain match confidence, suggest reconciliation rules \\
\hline
Model verification and logic enforcement & \textit{Alloy Analyzer} \shortcite{alloy2006}, HermiT Reasoner \shortcite{hermit2014}, SHACL validators & Converts informal constraints to formal rules; explains violations & LLM provides human-readable feedback on validation results; tools execute logic checks \\
\hline
Process modeling and simulation specification & \textit{Wolfram SystemModeler} \shortcite{systemmodeler2023}, MATLAB/Simulink, Cameo & Converts scenarios into system dynamics / process models & LLM prompts must be aligned with domain-specific syntax; tool ensures compile-ready output \\
\hline
Business rule formalization & \textit{Erwin} \shortcite{erwin2023}, \textit{Enterprise Architect} \shortcite{enterprise2023} & Converts high-level requirements to constraint logic & Tool checks completeness, syntax; LLM supports traceability explanation \\
\hline
Documentation and model explanation & OWL2NL, GPT-4 + LangChain \shortcite{langchain2023} & Verbalizes formal models; generates stakeholder reports & LLM expands sparse documentation; tool outputs serve as structured input \\
\hline
Workflow orchestration (multi-tool) & GPT-4 + Toolchain (via LangChain and API) & Dispatches subtasks to appropriate tools; sequences steps & Requires modular design; tools expose APIs or plugin interfaces \\
\hline
\end{tabular}
\label{tab:llm-integration-patterns}
\vspace{-0.9em} 
\captionsetup{skip=0pt}
\caption*{\footnotesize \textit{Note: RDF is the Resource Description Framework, SHACL is the Shapes Constraint Language for describing RDF graphs.}}
\end{table}

Model validation is natively supported by tools such as the Alloy Analyzer \shortcite{alloy2006}, which provides bounded model checking for first-order logic specifications, and HermiT \shortcite{hermit2014}, a widely-used reasoner for OWL ontologies. These tools help verify that structural constraints and logical rules are satisfied, reducing errors prior to simulation or implementation into broader workflows. Tools such as Enterprise Architect \shortcite{enterprise2023} offer automatic report generation from structured models, while OWL2NL translates ontology axioms into readable natural language. These features support stakeholder communication, requirements documentation, and training. Additionally, tasks such as data and schema modeling are directly supported out-of-the-box by tools such as ER/Studio and Erwin Data Modeler \shortcite{erwin2023}, both of which allow analysts to design ER or UML diagrams, enforce schema constraints, and integrate with enterprise databases. Lastly, system modeling and simulation is supported natively by Wolfram System Modeler \shortcite{systemmodeler2023}, which uses the Modelica language to represent complex multi-domain systems and provides built-in simulation capabilities—particularly useful in biomedical, engineering, and manufacturing applications.

In some cases, general-purpose modeling tools or libraries can accomplish identified  modeling for M\&S tasks with user-driven integration. Table 2 enumerates such scenarios. No single tool from the table can, in isolation, produce a validated, domain-specific  model; modelers must orchestrate multiple tools in sequence to that end. Prior to LLMs, such orchestration was already common: one might apply an NLP pipeline—e.g., using a library like spaCy—to extract candidate entities or relationships from text, then employ a schema matching system or data cleaning tool, such as OpenRefine \shortcite{openrefine2023}, to reconcile terminology, and finally use an ontology editor or modeling environment to formalize the concepts. LLM-integrated workflows build upon the same division of labor, where an LLM takes on a `glue' role: mediating between human language and the formalisms required by each specialized tool.

\begin{figure}
    \centering
    \includegraphics[width=0.55\linewidth]{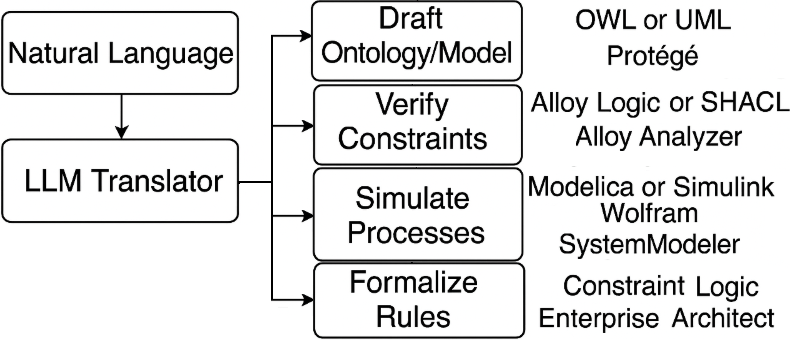}
    \caption{LLMs as Middleware for Modeling Tasks}
    \label{fig:workflow}
\end{figure}

While many core modeling tasks are supported by specialized tools out-of-the-box, numerous other tasks remain difficult to execute without integrating multiple tools or bridging semantic gaps between representations. These include cases where natural language must be translated into formal structure, where informal business rules must be formalized, or where model components from heterogeneous systems must be aligned. Here, LLMs can serve as a middleware, translating, sequencing, and coordinating modeling subtasks (Figure~\ref{fig:workflow}). Table 2 presents LLM-mediated  modeling workflows from initial model drafting to rule formalization and documentation, highlighting common subtasks, the target representation language(s) that the LLM must generate (e.g., OWL, UML), the specialized tool(s) capable of ingesting and executing that representation (e.g., Protégé), and the role of the LLM as a mediator between natural language and formal artifacts (e.g., translation, summarization). 

The LLM serves not as a substitute for the formal tool, but as an intelligent bridge--converting human intent into machine-verifiable representations (Figure~\ref{fig:workflow}). The table operationalizes this division of labor, offering practical guidance on how to scope LLM-mediated translation workflows depending on the modeling need. This directly addresses \textbf{Q1} by establishing clear, task-specific routes from language to tool.

For example, when drafting a model from natural-language requirements, an LLM such as GPT-4 can be used to generate structured output in OWL, UML, or ER formats, which can then be validated and refined in tools like Protégé, UModel, or ER/Studio. Here, the LLM is a translator that outputs standards-compliant syntax – such as RDF/XML - enabling handoff to verification-capable environments. In more complex tasks like ontology alignment or schema reconciliation, LLMs may again play a mediating role, revealing candidate terms for mappings, resolving ambiguity, or interpreting tool-generated error logs. Tools such as OntoAligner and OpenRefine remain central for alignment and cleanup, but LLM can offer contextual explanations and generate documentation. Related, when enforcing model constraints or validating logical coherence, LLMs may assist by transforming high-level rules into formal expressions suitable for execution in tools like Alloy Analyzer or HermiT. Here again, LLMs also help translate validation failures into actionable, human-readable documentation that reduces debugging time.

In process modeling and simulation workflows, LLMs may convert textual descriptions into system dynamics models in Modelica, block diagrams in Simulink, or behavior trees in Cameo. These conversions require structured prompting and domain-specific syntax, which the modeling tool ultimately compiles and validates. Overall, LLMs can orchestrate multi-step modeling workflows by dispatching tasks to different tools, coordinating formats, and sequencing outputs. This pattern is increasingly supported by frameworks like LangChain and toolchains that expose plugin APIs, allowing for modular LLM-driven pipelines.

The utility of LLMs as middleware is made more apparent when combining tasks. For example, if an LLM proposes a set of UML class definitions, a UML parser or modeling environment can be invoked to parse that proposal. If the parser returns errors (say, an undefined class reference or a syntax mistake in the UML), the error messages can be fed back to the LLM with a prompt to fix the issues. The LLM, now aware of where it went wrong, can adjust the class definitions accordingly. This iterative loop continues until the specialized tool confirms that the model is syntactically and structurally sound. Such a strategy leverages the LLM’s strength in rapidly generating and regenerating content, while the tool’s strictness guarantees eventual correctness. It is a form of algorithmic scaffolding, where the LLM is never left unguided for too long – the tool provides guardrails at each step.
The distinction between Table 1 and Table 2 underlines a recommendation: whenever a task can be done with a purpose-built tool, prefer to integrate the LLM with that tool, rather than relying on the LLM alone or reinventing logic via scripting. 

\section{LLMs for simulation}
After we discussed in the last section how LLMs can support the modeling process better, we focus in this section on the simulation development process.
\label{sec:simulation}


\subsection{Requirements analysis}

In the earliest phase of the M\&S lifecycle, artifacts are typically encoded in natural language (e.g., requirements and specifications), which opens some opportunities for LLMs.
LLMs can be used, e.g., to \textit{analyze project requirements} at a high level before commencing more resource-intensive formal specification and problem decomposition tasks. Such analyses can be used, e.g., to \textit{recommend appropriate simulation paradigms, formalisms, and tools} for the project. LLMs can rely on very diverse contextual data to drive such recommendations, including the comprehensive analysis of a corpus describing a problem, available data, and the skill set at hand.
Such use cases raise the question whether LLMs can be used to \textit{bypass  modeling} in the traditional requirements 
--  modeling -- simulation process.
Finally, LLMs can be used for validating requirements, e.g., by \textit{producing pseudocode}~\shortcite{xu2024pseudo}.

\subsection{Source code engineering}

The strong link between LLMs and software code is rooted in the well-recognized naturalness hypothesis of software~\shortcite{hindle2016naturalness}, which states that program code is a form of human communication bearing similar statistical properties to those of natural languages.
%
\textit{Source code generation} is among the key techniques in this lifecycle phase, and is has been a subject to thorough investigation.
\shortciteN{gerstmayr2024multibody} use LLMs to generate multi-body system dynamics models from natural language specifications.
\shortciteN{valeriomiceli-barone2023dialogue-based} translate user instructions to code for self-driving car simulations.
\shortciteN{liu2024languaging} generate executable simulation code from human language descriptions.
\shortcite{jackson2024natural} generate executable simulation for a logistic systems for inventory and process control.
\shortciteN{wang2024gensim} generate code for rich simulation environments and expert demonstrations for robot training.
\textit{Code translation and adaptation} between tools in simulation tool chains is another important case of source code engineering by LLMs.
This case is gaining attention in research. For example, \shortciteN{shrestha2021slgpt} use a GPT model to generate graphical block-diagram models, such as Simulink models, and pre-process the generated models for rapid search.
However, as shown by \shortciteN{pan2023understanding} through their experiments with 1\,700 code samples in C, C++, Go, Java, and Python, LLMs are not yet reliable enough for code translation, with the reported proportion of correct translations ranging between 2.1\% and 47.3\%.

\subsection{Testing and V\&V}
In the testing and V\&V lifecycle phase, LLMs are typically used for \textit{generating test cases and test artifacts} from simulation models or code, and usage examples.
%
%
Such avenues have been explored by \shortciteN{schafer2024empirical} who generate unit tests with a median statement coverage of above 70\% and branch coverage above 52\%. LLMs can also generate various tests artifacts for simulation testing. For example, \shortciteN{saiaswathduvvuru2025llm-agents} generate environmental configurations, test cases, and test properties (specific metrics, e.g., target detection and identification accuracy) for testing uncrewed aerial systems.
%
%
Another application of LLMs is \textit{verification}, when more rigorous evidence, e.g., formal proof is required to guarantee key system properties.
For example, \shortciteN{hassan2024llm-guided} use LLMs to generate formal specifications from natural language specifications, and to generate system invariants for a high-performance theorem prover (Z3) and summarize them in natural language.

\subsection{Documentation}
Documentation is a natural use case of LLMs, for the same reason code engineering is: due to the naturalness hypothesis of software~\shortcite{hindle2016naturalness}. Since source code exhibits the same statistical traits as human language, it is rather straightforward to employ LLMs for \textit{generating human-readable explanations} of source code, simulation models, and simulation results at various levels of technical depth. An example is given by \citeN{jackson2023natural}.
Similar avenues have been explored, e.g., by \shortciteN{nam2024understanding} to generate comprehensive documentation for complex simulations.
A special form of documentation is \textit{summarization} of source code and simulation models that aims to produce brief and target human-readable descriptions for experts. For example, \shortciteN{ahmed2023few-shot} use GPT models with few-shot learning to summarize source code, and their evidence suggests that LLMs significantly surpass other state-of-the-art models for code summarization. Similarly, \shortciteN{khan2023automatic} generate documentation using a GPT-3 based model, Codex, pre-trained on both natural and programming languages, and achieve a BiLingual Evaluation Understudy (BLEU) score over 0.2, a 11\% improvement over state-of-the-art techniques.

\section{Architectures to Efficiently Integrate LLMs as M\&S Middleware}
\label{sec:architectures}

\begin{figure}[h]
    \centering
    \includegraphics[width=\textwidth]{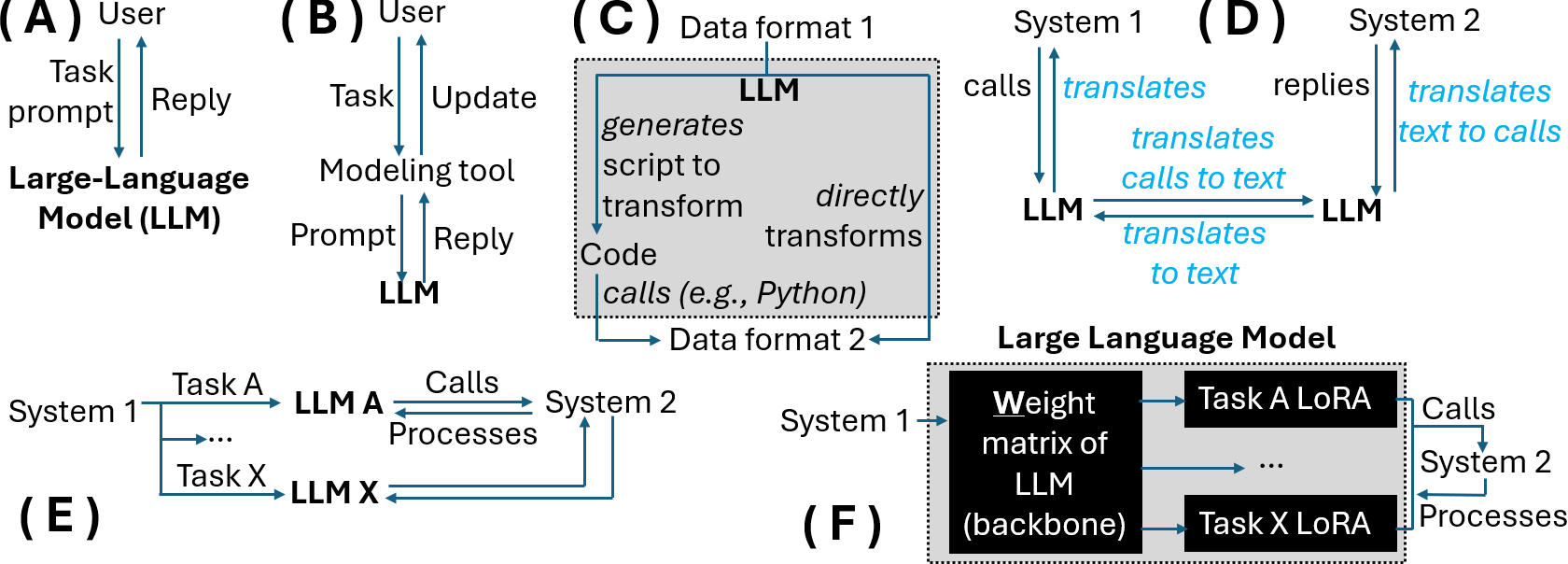}
    \caption{The simplest use of LLMs is to set a task and use the reply either directly (A) or via a software (B), but a LLM may generate a script instead of directly solving a task itself (C) or a set of LLMs may support system interoperability (D--E). In contrast with these prior frameworks, we recommend handling for M\&S tasks with one LLM to improve computational efficiency and maintainability (F).}
    \label{fig:architectures}
\end{figure}

A simple use of LLMs is to prompt them with a task and directly use their answer (Figure~\ref{fig:architectures}-A), possibly through a follow-up prompt. However, as the use of LLMs within M\&S matured, we observe a shift towards training and refining LLMs for specific tasks and/or integrating LLMs with other tools through a workflow. As a case in point, researchers first used GPT through prompts to turn a prose-based narrative into simulation code~\shortcite{frydenlund2024modeler}, and later ~\shortciteN{martinez2024enhancing} fine-tuned the system (few-shot prompting) and connected it to a database (retrieval-augmented generation). Newer architectures reflect these changes in practices (Figure~\ref{fig:architectures}-B), for instance when a modeling tool such as a class diagram editor internally prompts the LLM to recommend classes or attributes~\shortcite{chaaben2024utility}. This section thus examines  approaches that employ LLMs as a middleware, emphasizing support for M\&S tasks.



There is also the potential to use LLMs as translators between two different representations of knowledge. While several studies bridged natural language and SQL commands for a database~\shortcite{LLM-DB124,LLM-DB224}, mediating between different schema has received less attention. Developing these methods further can lead to the automatic configuration of data mediation services between different structures of knowledge representation while ensuring their conceptual alignment~\cite{tolk2024conceptual}. \shortciteN{calvetcan} compared two architectures to use LLMs as data conversion tools, possibly within a single system. They noted that prior works performed a \textit{direct} conversion: users provide one data format to the LLM, which must generate data in the target format. The authors introduced an \textit{indirect} step by generating Python code and calling it to execute the conversion (Figure~\ref{fig:architectures}-C). The advantage is subtle yet important: a direct conversion may work once but there is no guarantee that it will work next time, due to stochasticity in the LLMs. In contrast, if a Python code is correctly generated once, then it can be used to correctly perform all data conversions. The authors also showed that LLM-based translations can be challenging, as the relatively simple data schema used in the study could not always be fully converted and some LLMs were never able to generate Python code to perform the conversion. The takeways are twofold: (i) transforming a problem instance into a target domain language and executing it can improve system reliability compared to performing all tasks via the LLM, but (ii) some transformations can require significant engineering efforts.

\citeN{lehmann2024towards} proposed an architecture consisting of two LLMs, one for the `consumer' and another for the `provider' (Figure~\ref{fig:architectures}-D). His vision was unique in using natural text as intermediate between LLMs: the consumer LLM does not \textit{directly} pass calls onto the target system or the provider LLM; rather, it translates the call into text, which is translated back into a call by the provider LLM. As a result, this architecture involves four translations. The author noted the cumulative risks posed by these multiple translations, ``adding multiple components into the communication between consumer and provider also introduces potential points of failure'', thus we do not retain this architecture for M\&S applications.

The `pretrain-then-finetune' paradigm results in a multiplicity of specialized LLMs, each with (sometimes minor) differences in the weights of their neural network. As a result, if a modeler switches from task A to task B, they need to unload the LLM for task A and load the LLM for task B (Figure~\ref{fig:architectures}-E). In application settings where data is sensitive and computations must be performed locally (e.g., enterprise modeling, healthcare, defense), repeatedly unloading and loading a locally deployed LLM becomes a performance bottleneck as it is taxing on memory and takes time. In contrast, a parameter-efficient fine-tuning method such as Low-Rank Adaptation (LoRA) introduces small (low-rank) matrices into the weight updates of a pre-trained model, rather than fine-tuning all parameters~\shortcite{yang2024low}. As a result, the bulky LLM is loaded only once, and light-weight matrices build on this shared backbone to support specific tasks (Figure~\ref{fig:architectures}-F). This avoids the taxing loading-and-unloading steps for memory management~\cite{chen2024comparative}, and it also improves the training process since training a LoRA is significantly faster and more efficient than fully fine-tuning a large neural network -- computations can be further improved through approximated LoRA~\shortcite{chen2025lora}. Several systems can manage LoRA adapters on a shared backbone and route user requests along different task-specific LoRAs, from the early {\ttfamily Punica}~\shortcite{chen2024punica} to the recent {\ttfamily dLoRA}~\shortcite{dou2024loramoe}.

We note that this research area is still in its infancy, thus there is ongoing work on integrating adapters for optimal performance. In particular,~\shortciteN{tian2025adapters} proposed creating a `middleman adapter' (or smart routing system) that identifies a task from the user input and subsequently matches the task with the right LoRA. It is thus \textit{essential for the M\&S field to develop high-quality specialized LoRAs and automatically map them onto tasks}. The literature on M\&S building blocks suggests that it is preferable to identify the tasks faced by modelers and then create the corresponding LoRAs~\shortcite{cheng2023identifying,schroeder2022towards}, rather than creating specialized systems and then trying to map them to user needs. We thus recommend a comprehensive analysis of modeling tasks (i.e., user needs gathering), culminating in the generation of LoRAs covering several orthogonal needs. We posit that such a bottom-up process would also facilitate the alignment of user prompts and their correct routing through a LoRA.

\section{Discussion}
\label{sec:discussion}

\subsection{Model transformation}

Trusting LLMs as translators between systems depends on the definition of `trust' and the specific context in which they are used. If trust is defined as \textit{absolute reliability} with zero distortion or hallucination, then LLMs cannot be fully trusted, as they can introduce subtle errors or fabrications. However, trust can be framed in a \textit{comparative sense} by evaluating the distortion caused by LLMs against existing solutions. For instance,~\shortciteN{yoon2024redefining} converted data back-and-forth between two formats, with an LLM and with traditional rule-based methods. In their context of health data, the LLM distorted and lost data less than the rule-based methods currently in use. In addition to measure the amount of errors, it is useful to understand the nature of these errors. Errors do not all have the same consequences and translators do not all have the same predictability: if a rule-based system tends to misclassify certain types of entries then we \textit{know} the errors, whereas hallucinations from LLMs may not be as predictable. Ultimately, trust should not be considered as binary but as a balance of accuracy, predictability, and the ability to mitigate errors through oversight and verification.

This paper is an important step in identifying potential tools and languages for each step encountered in the M\&S process, but these choices have an impact.~\shortciteN{lam2024closer} reported a near 50\% performance variation when the LLM translated between different representations and employed various tools to determine whether a conclusion followed from a set of premises. It is thus necessary to extend emerging M\&S benchmarks~\shortcite{giabbanelli2025benchmarking} to account for the choice of tools and representations, such that modelers identify the best option available in their context. Designing benchmarks is challenging, as even small elements such as fictional character names may impact the LLM's performance~\shortcite{saparov2023testing}.

In addition, training LLMs to support multiple translations (e.g., via plethora of LoRAs) may become cumbersome, particularly if we seek to certify these translations or maintain their accuracy in the face of evolving language specifications. Rather than envisioning that all translations \textit{directly} produce the language of interest, it may be more pragmatic to consider \textit{indirect} or multi-hops translations. That is, the M\&S community may maintain a core set of translations (e.g., if the user's query calls for first-order logic then use {\ttfamily Prover9} syntax) and leave it to users if they need minor adjustments afterward (e.g., turn {\ttfamily Prover9} into the related {\ttfamily Pyke} syntax to use a different solver). As previously noted, supporting interoperability by including indirect translations may lead to different sequences (e.g., turn task description into $Z$ then $C$ or $A$, $B$ then $C$), which may have different computational costs or accuracies~\shortcite{schuerkamp2023facilitating}. Such problems have been studied in multi-paradigm modeling (MPM)~\shortcite{vangheluwe2002introduction} using \textit{formalism-transformation graphs} (FTGs) that represent formalisms as nodes of a graph and transformations as edges~\shortcite{challenger2020ftgpm}. A process model that instantiates formalisms as models, and transformations as activities, allows for the thorough investigation of complex transformation chains~\shortcite{lucio2013ftgpm}.

An approach worth following is the combination of ontological methods and LLMs, such as exemplified in \shortcite{LLM-Onto24}. The underlying idea is to use ontologies as the formal specification of the conceptualization of the information to be shared as well as its structure. LLM expression must fall into this specification to be producible for the specified sending systems and vice versa to be understandable by the receiving system. Using the ontological specification of sender and receiver avoids the acceptance of hallucinations outside of the valid range of scope and structure while benefiting from the convenience of using the LLM. As demonstrated by~\citeN{jackson2023natural}, LLMs can describe even simulation systems in natural language, so that the description of interfaces and their constraints should be supported similarly. However, this approach is still in its early stages.

While our paper focused on the use of LLMs as middleware, it does not mean that LLMs merely act as messengers that translate requests between users and specialized tools. On the one hand, the tools can improve the LLM by amplifying its capabilities (e.g., a solver ensures that reasoning is coherent) or giving it feedback (e.g., errors reported by tools can be integrated in error-solving mechanisms as in~\shortciteNP{pan2023logic}). On the other hand, \textit{the LLM may help the tools}. For instance, the user may have a query in which some conditions are so obvious that they are not explicitly stated. A symbolic solver by itself may struggle when it expects a full chain of logic, but the LLM may provide the missing information. The architectures envisioned here are thus a simplification of the exciting potential mutual benefits that remain to be explored between LLMs and tools.

\footnotesize

\bibliographystyle{wsc}
\bibliography{references}

\end{document}